\documentclass[pra,twocolumn,showpacs,superscriptaddress,amsfonts,amsmath,floatfix]{revtex4}

\usepackage{graphicx}

\begin{document}

\title{Free expansion of a Lieb-Liniger gas:
Asymptotic form of the wave functions}

\author{D.~Juki\' c}
\affiliation{Department of Physics, University of Zagreb, PP 332, 10000 Zagreb, Croatia}
\author{R.~Pezer}
\affiliation{Faculty of Metallurgy, University of Zagreb,
Aleja narodnih heroja 3, 44103 Sisak, Croatia}
\author{T.~Gasenzer}
\affiliation{Institut f\"ur Theoretische Physik,
Universit\"at Heidelberg, Philosophenweg 16, 69120 Heidelberg, Germany}
\affiliation{Kavli Institute for Theoretical Physics,
University of California, Santa Barbara, CA 93106-4030}
\author{H.~Buljan}
\email{hbuljan@phy.hr}
\affiliation{Department of Physics, University of Zagreb, PP 332, 10000 Zagreb, Croatia}

\date{\today}

\begin{abstract}
The asymptotic form of the wave functions describing
a freely expanding Lieb-Liniger gas is derived by using a
Fermi-Bose transformation for time-dependent states,
and the stationary phase approximation.
We find that asymptotically the wave functions approach the 
Tonks-Girardeau (TG) structure as they vanish when any two of the particle 
coordinates coincide. 
We point out that the properties of these asymptotic states 
can significantly differ from the properties of a TG gas 
in a ground state of an external potential. 
The dependence of the asymptotic wave function on the initial 
state is discussed. 
The analysis encompasses a large class of initial conditions,
including the ground states of a Lieb-Liniger gas in physically
realistic external potentials.
It is also demonstrated that the interaction energy asymptotically 
decays as a universal power law with time, $E_\mathrm{int}\propto t^{-3}$.

\hfill HD--THEP--08--10\\[-5ex]
\end{abstract}

\pacs{05.30.-d, 03.75.Kk, 67.85.De}
\maketitle

\section{Introduction}

The physics of one-dimensional (1D) Bose gases in many aspects
differs from the physics encountered in higher dimensional systems.
For example, the Lieb-Liniger (LL) gas of $\delta$-interacting
bosons in one spatial dimension
becomes less ideal as its density decreases
\cite{Lieb1963}, and eventually approaches the Tonks-Girardeau (TG)
limit of a gas of "impenetrable-core" bosons \cite{Girardeau1960} as it
becomes sufficiently diluted. The interest in these 1D systems
is greatly stimulated by their experimental realization with
atoms confined in tight 1D atomic wave guides
\cite{OneD,TG2004,Kinoshita2006}. The special features of
effectively 1D atomic gases \cite{Olshanii,Petrov,Dunjko} are
reflected by properties of nonequilibrium dynamics in
these systems, which have become accessible experimentally
\cite{Kinoshita2006}. The possibility of finding exact
time-dependent solutions for LL
\cite{Gaudin1983,Girardeau2003,Buljan2008} and TG
\cite{Girardeau2000,Girardeau2000a,Ohberg2002,Busch2003,
Rigol2005,Minguzzi2005,DelCampo2006,Rigol2006,Buljan2006,Pezer2007,
Gangardt2007} evolution
is of particular theoretical interest, as they have the
potential to provide insight beyond various approximation
schemes.

Exact solutions for a homogeneous Bose gas with (repulsive)
point-like interactions of arbitrary
strength $c$, and periodic boundary conditions,
were presented by Lieb and Liniger in 1963 \cite{Lieb1963}.
For attractive interactions, $c<0$, exact LL wave functions were analyzed
in Ref.~\cite{McGuire1964}. The case
of box confinement for $c>0$ was studied in Ref.~\cite{Gaudin1971}.
In the light of recent experiments \cite{OneD,TG2004,Kinoshita2006}
exact studies of the LL model are even more attractive today
\cite{Muga1998,Sakmann2005,Batchelor2005,Kinezi2006,Sykes2007}.
Besides providing insight into the physics of 1D Bose
gases, exact solutions can serve
as a benchmark for various approximations as well as for
numerical approaches (see, e.g., Refs.~\cite{Sykes2007,Kanamoto2005}).
The calculation of correlation functions of a LL gas from the
wave functions is a difficult task; these functions
furnish observables like the momentum distribution of particles in the
gas, and were studied by using various approaches
(e.g., see Refs.~\cite{Creamer1981,Jimbo1981,Korepin1993,Kojima1997,Olshanii2003,Gangardt2003,
Astrakharchik2003,Kheruntsyan2005,Forrester2006,Caux2007,Calabrese2007,Khodas2007}).
Time-dependent phenomena in the context of LL gases
with finite-strength interactions have been addressed by using
both analytical \cite{Gaudin1983,Girardeau2003,Buljan2008}
and numerical methods (see, e.g., Refs.~\cite{Berman2004,Li2005}).
Irregular dynamics of a LL gas was studied numerically in
a mesoscopic system in Ref. \cite{Berman2004}.
In Ref.~\cite{Girardeau2003}, it was shown that phase imprinting
by light pulses conserves the so-called cusp condition for the LL wave function
imposed by the interactions.

Exact solutions for 1D Bose gases are conveniently constructed
by using the Fermi-Bose mapping techniques
\cite{Girardeau1960,Gaudin1983,Girardeau2000,Das2002}.
In 1960 Girardeau discovered that the wave function of a
spinless noninteracting 1D Fermi gas can be symmetrized such
that it describes an impenetrable-core 1D Bose gas
\cite{Girardeau1960}. This mapping is valid for arbitrary
external potentials \cite{Girardeau1960}, for time-dependent
problems \cite{Girardeau2000}, and in the context of
statistical mechanics \cite{Das2002}.
In fact, fermion-boson duality in 1D exists for arbitrary interaction strengths
\cite{Cheon1999,Yukalov2005}.
Furthermore, a time-dependent antisymmetric
wave function describing a 1D system of noninteracting fermions can be
transformed, by using a differential Fermi-Bose mapping operator, to
an exact time-dependent solution for a LL gas, as outlined by Gaudin~\cite{Gaudin1983}.
This method is applicable in the
absence of external potentials and other boundary conditions.
Therefore, it is particularly useful to study free expansion of
LL gases from an initially localized state.

Free expansion of interacting Bose gases has recently attracted considerable
attention. It has been utilized in
experiments to deduce information on the initial state (see, e.g.,
Ref.~\cite{Bloch2008} and references therein), and can be considered as a
quantum-quench-type problem which provides insight into the relaxation
of quantum systems (see, e.g., Refs.~\cite{Calabrese2007a,Rigol2007a}
and references therein).
Free expansion of a LL gas has been analyzed in Ref. \cite{Ohberg2002} 
by employing the hydrodynamic formalism \cite{Dunjko}; it was shown that the 
density of the gas does not follow self-similar evolution \cite{Ohberg2002}. 
However, in 1D Bose systems, most exact many-body solutions are given for the 
TG gas \cite{Ohberg2002,Rigol2005,Minguzzi2005,DelCampo2006,Gangardt2007}.
An important result is that the momentum distribution
of the freely expanding TG gas asymptotically approaches
the momentum distribution of free fermions \cite{Rigol2005,Minguzzi2005}.
Recently, we have constructed a particular family
of exact solutions describing a LL gas freely expanding from a
localized initial density distribution \cite{Buljan2008}.
It was shown that for any interaction strength, the wave functions
asymptotically (as $t\rightarrow \infty$) assume TG form.
Even though it is generally accepted that 1D
Bose gases become less ideal with decreasing density,
this intuition is mainly based on the studies of a LL gas 
in equilibrium ground states \cite{Lieb1963}. 
Thus, a more rigorous analysis of the expanding LL gas, which 
leads to more dilute system, but out of equilibrium, is desirable. 
In particular, it is interesting to study the dependence of the 
asymptotic wave functions on the initial state, and to see how are the 
initial conditions imprinted in the asymptotic states.

Here we study the asymptotic form of the wave function
describing a freely expanding Lieb-Liniger gas,
which can be constructed via the Fermi-Bose transformation
and the stationary phase approximation.
In Section \ref{sec:LL} we describe the LL model and the
Fermi-Bose transformation.
In Section \ref{sec:ExpAsymp} we demonstrate that the asymptotic
wave functions have Tonks-Girardeau structure,
that is, they vanish when any of the two particle coordinates coincide. 
The dependence of the asymptotic state on the initial state is discussed. 
We illustrate that the properties of the asymptotic wave functions 
can significantly differ from the properties of a TG gas in the 
ground state of some external potential. 
This study generalizes and adds upon our previous result from Ref.~\cite{Buljan2008},
as the initial conditions studied here encompass ground states
for generic external potentials and various interaction strengths.
From the next-to-leading order term in the asymptotic regime,
we deduce that the interaction energy of the LL gas decays
as a universal power law in time $E_\mathrm{int}\propto t^{-3}$.
This is illustrated on a particular example in Section
\ref{Sec:Check}, where we provide further analysis of the particular family of 
time-dependent LL wave functions studied in Ref.~\cite{Buljan2008}.
Explicit expressions for the asymptotic form of the
single-particle density are provided in Section \ref{Sec:SPden}.
In Section \ref{Sec:Hydro} we calculate the asymptotic single-particle
density for free expansion of a LL gas from an infinitely deep box 
potential. We compare our exact calculation with the hydrodynamic approximation 
introduced in Ref. \cite{Dunjko}, and employed in Ref. \cite{Ohberg2002}
in the context of free expansion, obtaining good agreement 
for all values of the interaction strength.

\section{The Lieb-Liniger model}
\label{sec:LL}

A system of $N$ identical $\delta$-interacting bosons in
one spatial dimension is described
by the many-body Schr\"odinger equation \cite{Lieb1963}
\begin{equation}
i \frac{\partial \psi_B}{\partial t}=
-\sum_{i=1}^{N}\frac{\partial^2 \psi_B}{\partial x_i^2}+
\sum_{1\leq i < j \leq N} 2c\,\delta(x_i-x_j)\psi_B.
\label{LLmodel}
\end{equation}
Here, $\psi_B(x_1,\ldots,x_N,t)$ is the time-dependent wave function,
and $c$ is the strength of the interaction.
It is assumed that the initial wave function is localized,
e.g., by the system being trapped within some external potential,
before, at $t=0$, the trap is suddenly switched off and the gas starts
expanding.
We are interested in the behavior of $\psi_B$ when $t\rightarrow\infty$.
Here the spatial dimension is infinite
$x_j \in(-\infty,\infty)$, i.e.,
we do not impose any boundary conditions.

Due to the Bose symmetry of the wave function,
it is sufficient to express it in the fundamental sector of the
configuration space, $R_1:x_1<x_2<\ldots<x_N$, where
$\psi_B$ obeys
\begin{equation}
i \frac{\partial \psi_B}{\partial t}=
-\sum_{i=1}^{N}\frac{\partial^2 \psi_B}{\partial x_i^2}.
\label{free}
\end{equation}
The $\delta$-interactions
create a cusp in the wave function when two particles touch.
This can be expressed as a boundary
condition at the borders of $R_1$ \cite{Lieb1963}:
\begin{equation}
\left [
1-\frac{1}{c}
\left (
\frac{\partial}{\partial x_{j+1}}-\frac{\partial}{\partial x_j}
\right)
\right]_{x_{j+1}=x_j}\psi_B=0.
\label{interactions}
\end{equation}
These boundary conditions can easily be rewritten for any
permutation sector.
In the TG limit, i.e., for $c\rightarrow \infty$, the cusp
condition implies that the wave function vanishes when
two particles are in contact:
$\psi_B(x_1,\ldots,x_j,x_{j+1},\ldots,x_N,t)|_{x_{j+1}=x_j}=0$
\cite{Girardeau1960,Girardeau2000}.

Exact solutions of the time-dependent Schr\"odinger equation (\ref{LLmodel})
can be obtained by using
a Fermi-Bose mapping operator \cite{Gaudin1983,Buljan2008} acting on
fermionic wave functions:
If $\psi_F(x_1,\ldots,x_N,t)$ is an antisymmetric (fermionic)
wave function, which obeys the Schr\" odinger equation
for a noninteracting Fermi gas,
\begin{equation}
i \frac{\partial \psi_F}{\partial t}=
-\sum_{i=1}^{N}\frac{\partial^2 \psi_F}{\partial x_i^2},
\label{SchF}
\end{equation}
then the wave function
\begin{equation}
\psi_{B,c}= {\mathcal N}_{c} \hat O_c \psi_F,
\label{ansatz}
\end{equation}
where
\begin{equation}
\hat O_c=\prod_{1\leq i < j \leq N}
\left[
\mbox{sgn}(x_j-x_i)+\frac{1}{c}
\left(
\frac{\partial}{\partial x_{j}}-
\frac{\partial}{\partial x_{i}}
\right)
\right],
\label{oO}
\end{equation}
is the differential Fermi-Bose mapping operator, and
${\mathcal N}_{c}$ is a normalization constant,
obeys Eq.~(\ref{LLmodel}) \cite{Gaudin1983}. For the purpose of completeness
we outline, in Appendix \ref{app:FB}, the proof that
the wave function (\ref{ansatz}) obeys both the cusp condition imposed
by the interactions and the Schr\"odinger equation (\ref{free}).

\section{Free expansion: Asymptotics}
\label{sec:ExpAsymp}
In this section we study the asymptotic form of
time-dependent LL wave functions $\psi_{B,c}$
which are obtained by the Fermi-Bose transformation
(\ref{ansatz}).
All information on the initial condition $\psi_{B,c}(x_1,\ldots,x_N,t=0)$
is contained in the initial fermionic wave function
$\psi_F(x_1,\ldots,x_N,t=0)$:
\begin{equation}
\psi_{B,c}(x_1,\ldots,x_N,0)=
{\mathcal N}_{c} \hat O_c \psi_F(x_1,\ldots,x_N,0).
\label{initial}
\end{equation}
The initial bosonic wave function, which
can be expressed in this way, is assumed to describe
a LL gas in its ground state when trapped in some external potential $V(x)$,
e.g., in a harmonic oscillator potential, or some other trapping
potential used in experiments.
We consider the evolution from this initial
state after the trapping potential has been
suddenly turned off, as studied in experiments to
deduce information on the initial state \cite{Bloch2008}.
The time-dependent fermionic wave function $\psi_F(x_1,\ldots,x_N,t)$,
which freely expands from the initial condition $\psi_F(x_1,\ldots,x_N,0)$,
can be expressed in terms of its Fourier transform,
\begin{align}
&\psi_F(x_1,\ldots,x_N,t)
=\int dk_1 \cdots dk_N
\nonumber \\
&\quad\times\ \tilde \psi_F(k_1,\ldots,k_N)
e^{i \sum_{j=1}^{N} [k_j x_j - \omega(k_j) t]},
\label{psiFt}
\end{align}
where $\omega(k)=k^2$, and
\begin{align}
&\tilde \psi_F(k_1,\ldots,k_N)
=\frac{1}{(2\pi)^N}
\int dx_1 \cdots dx_N
\nonumber \\
&\quad\times\ \psi_F(x_1,\ldots,x_N,t=0)
e^{-i \sum_{j=1}^{N} k_j x_j}.
\end{align}
By using the Fermi-Bose transformation, the time-dependent
bosonic wave function describing the freely expanding LL gas
can be expressed as
\begin{align}
&\psi_{B,c}=\int dk_1 \cdots dk_N
\nonumber \\
&\quad\times\ G(k_1,\ldots,k_N)
e^{i \sum_{j=1}^{N} [k_j x_j - \omega(k_j) t]},
\label{psiBt}
\end{align}
where the function $G(k_1,\ldots,k_N)$ is defined as
\begin{align}
&G(k_1,\ldots,k_N) = {\mathcal N}_{c} \tilde \psi_F(k_1,\ldots,k_N)
\nonumber \\
&\quad\times\
\prod_{1\leq i<j\leq N}
[\mbox{sgn}(x_j-x_i)+\frac{i}{c}(k_j-k_i)].
\label{gk}
\end{align}
It should be noted that $G(k_1,\ldots,k_N)$ is {\em not}
the Fourier transform of $\psi_{B,c}$ because it depends
on $x_j$ through the $\mbox{sgn}(x_j-x_i)$ terms.

The asymptotic form of the wave function (\ref{psiBt}) can be
obtained by evaluating the integral with the stationary phase
approximation. The phase $\phi=\sum_{j=1}^{N} [k_j x_j -
\omega(k_j) t]$ is stationary when $\partial \phi/\partial
k_j=0$. Let $\{ k^{'}_j \}$ denote the $k_j$-values for which
\[
\left.\frac{\partial \phi}{\partial k_j}
\right|_{k^{'}_j}
=x_j - 2 k^{'}_j t=0,
\]
that is, $k^{'}_j=x_j / 2t$.
The phase can be rewritten as
\[
\phi(\{k\})=\phi(\{k^{'}\})-t \sum_{j=1}^N (k_j-k^{'}_j)^2.
\]
The leading term of the integral in Eq.~(\ref{psiBt}),
as well as the next-to-leading term, can be evaluated by
expanding $G(k_1,\ldots,k_N)\equiv G(\{k\})$ in a Taylor
series around the stationary phase point $\{k^{'}\}$:
\begin{widetext}
\begin{align}
\psi_{B,c}\ =\
&  e^{i \phi(\{k^{'}\})} \Big[ G(\{k^{'}\})
\int dk_1 \cdots dk_N e^{-it \sum_{j=1}^N (k_j-k^{'}_j)^2}
\nonumber \\
& +\ \sum_{i=1}^N \left.\frac{\partial G(\{k\})}{\partial k_i}\right|_{\{k^{'}\}}
\int dk_1 \cdots dk_N (k_i-k^{'}_i) e^{-it \sum_{j=1}^N (k_j-k^{'}_j)^2}
\nonumber \\
& +\ \frac{1}{2!} \sum_{i,j=1}^{N}
\left.\frac{\partial^2 G(\{k\})}{\partial k_i \partial k_j}\right|_{\{k^{'}\}}
\int dk_1 \cdots dk_N (k_i-k^{'}_i)(k_j-k^{'}_j)
e^{-it \sum_{l=1}^N (k_l-k^{'}_l)^2}\ +\ldots\Big].
\label{psiBtSeries}
\end{align}
\end{widetext}
The remaining integrals in the three terms written out in this expansion
can be calculated analytically.
The second term involving the first derivatives of $G(\{k\})$ vanishes.
The third term is nonvanishing only for $i=j$.
Thus Eq.~(\ref{psiBtSeries}) reduces to
\begin{align}
\psi_{B,c}
& =  e^{i \phi(\{k^{'}\})} \left(\sqrt{\frac{\pi}{t}}e^{-i\pi/4}\right)^N
\nonumber\\
&\qquad\Big[G(\{k^{'}\})-\frac{i}{4t}
\sum_{i=1}^{N}
\left.\frac{\partial^2 G(\{k\})}{\partial k^2_i}\right|_{\{k^{'}\}}
+\ldots\Big].
\label{psiBtSeries2}
\end{align}
From Eq. (\ref{psiBtSeries2}) we obtain in leading order the
asymptotic wave function
\begin{align}
\psi_{\infty} &\propto\ t^{-{N}/{2}}
\prod_{1\leq i<j\leq N}
\Big[\mathrm{sgn}(x_j-x_i)+\frac{i}{c}(k^{'}_j-k^{'}_i)\Big]
\nonumber \\
&\quad \times\ \tilde \psi_F(k^{'}_1,\ldots,k^{'}_N)
e^{i \sum_{j=1}^{N} [k^{'}_j x_j - \omega(k^{'}_j) t]},
\label{psiAsym1}
\end{align}
which is written in a more convenient form in terms of the
variables $\xi_j=x_j/t$:
\begin{align}
\psi_{\infty}
& \propto\ t^{-{N}/{2}}
\prod_{1\leq i<j\leq N}
\Big[\mathrm{sgn}(\xi_j-\xi_i)+\frac{i}{2c}(\xi_j-\xi_i)\Big]
\nonumber \\
&\quad\times\ \tilde \psi_F(\xi_1/2,\ldots,\xi_N/2)\,
e^{\frac{i}{4}\sum_{j=1}^{N} \xi_j^2 t}.
\label{psiAsym}
\end{align}
Equation (\ref{psiAsym}) is the main result of this paper.
Evidently the asymptotic form of the LL wave function
$\psi_{\infty}$ has TG form. Namely, the Fourier transform
of a fermionic wave function $\tilde \psi_F(\xi_1/2,\ldots,\xi_N/2)$
is antisymmetric, which implies that $\psi_{\infty}$ is zero
whenever $\xi_i=\xi_j$ $(i\neq j)$.
Furthermore, $\psi_{\infty}$ is symmetric under the exchange
of any two coordinates $\xi_i$ and $\xi_j$.
This clearly shows that a localized LL wave function 
during free expansion asymptotically approaches a wave function 
with the TG structure. However, it should be emphasized 
that the properties of the asymptotic state are not necessarily 
similar to the wave function describing TG gas in equilibrium, 
in the ground state of some external potential.
The connection between the initial and the asymptotic state
is illustrated below.

In the derivation of Eq.~(\ref{psiAsym}) we have analyzed
LL wave functions which are obtained through the
Fermi-Bose transformation (\ref{ansatz}). This class of
wave functions is quite general and corresponds to
numerous situations of practical relevance. Let us discuss
the case in which the initial bosonic wave
function $\psi_{B0}=\psi_{B,c}(x_1,\ldots,x_N,0)$ is
a ground state of a repulsive LL gas in an experimentally realistic
external potential $V(x)$, e.g., a harmonic oscillator potential.
The eigenstates of the LL system in free space are of the form
\begin{equation}
\psi_{\{ k \}}= \mathcal{N}({\{ k \}}) \hat O_c \det[e^{i k_{m}x_j}]_{m,j=1}^{N},
\label{expansionLL}
\end{equation}
where the set of $N$ real values $\{ k \}=\{ k_m\, |\, m=1,\ldots,N
\}$ uniquely determines the eigenstate; the normalization constant 
is given by
\[
\frac{1}{\mathcal{N}({\{ k \}})}=
\sqrt{ (2 \pi)^N N! \prod_{i<j}{\left[1+\left(\frac{k_j-k_i}{c}\right)^2 \right]}},
\] 
see Ref. \cite{Korepin1993}. 
In free space, there
are no restrictions on the numbers $k_m$. If periodic boundary
conditions are imposed as in Ref.~\cite{Lieb1963} (i.e., the
system is a ring of length $L$), the wave numbers $k_j$ must
obey a set of coupled transcendental equations
\cite{Lieb1963,Muga1998,Sakmann2005,Forrester2006,Sykes2007}
which depend on the strength of the interaction
(see, e.g., Ref.~\cite{Sakmann2005}).
The LL eigenstates $\psi_{\{ k \}}$ possess the closure property
\cite{Gaudin1983} and they are complete \cite{Dorlas1993}. 
Thus, our initial state $\psi_{B0}$ can be expressed as a
superposition of LL eigenstates,
\begin{eqnarray}
\psi_{B0} & = &
\sum_{\{ k \}} b(\{ k \}) \psi_{\{ k \}} \nonumber \\
& = & \hat O_c \sum_{\{ k \}} 
\mathcal{N}({\{ k \}}) b(\{ k \}) \det[e^{i k_{m}x_j}]_{m,j=1}^{N},
\label{LLcomplete}
\end{eqnarray}
where the coefficients $b(\{ k \})$ can be obtained by projecting
the initial condition $\psi_{B0}$ onto the LL eigenstates.
By comparing Eqs. (\ref{initial}) and (\ref{LLcomplete}) we find that 
the initial fermionic wave function is
\begin{equation}
\psi_{F0}= {\mathcal N}_{c}^{-1} \sum_{\{ k \}} 
\mathcal{N}({\{ k \}}) b(\{ k \}) \det[e^{i k_{m}x_j}]_{m,j=1}^{N}.
\label{psiF0viab}
\end{equation}
Since we have assumed that $V(x)$ is an experimentally realistic
smooth function, also $\psi_{F0}$ is smooth and differentiable
such that the operator $\hat O_c$ can be applied.
%
\begin{figure}
\begin{center}
\includegraphics[width=0.5 \textwidth ]{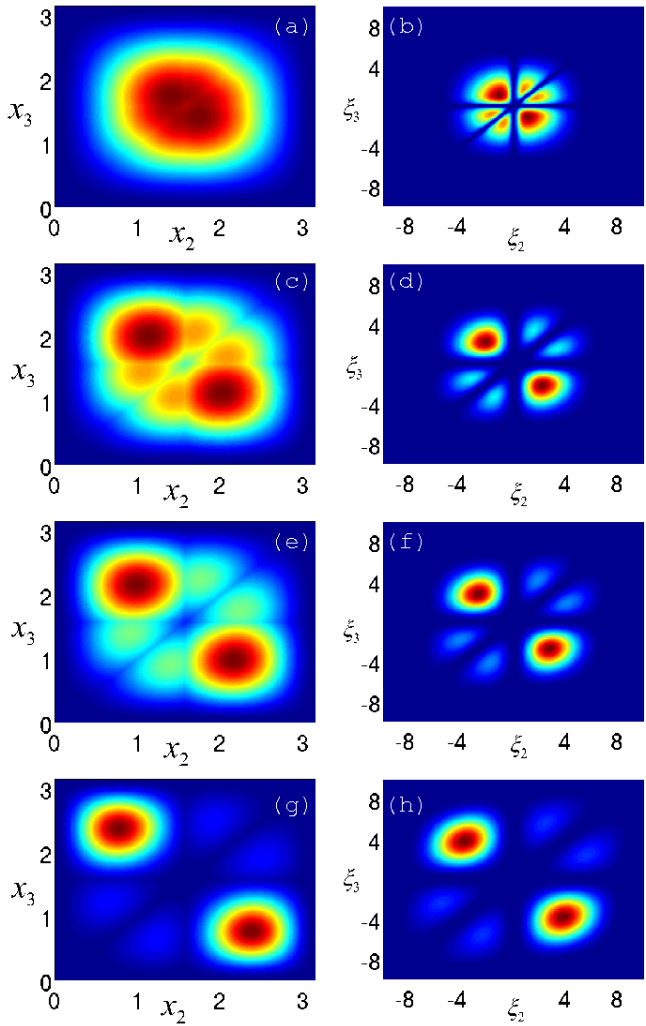}
\caption{ \label{figdensity}
(color online) Contour plots illustrating free expansion 
of $N=3$ bosons from the ground state of a LL gas in a box 
with infinitely high walls ($L=\pi$). 
The left column depicts the initial ground state $|\psi_{B0}(L/2,x_2,x_3)|^2$, and 
the right column depicts the asymptotic state $|\psi_{\infty}(0,\xi_2,\xi_3)|^2$, 
for $c=0.2$ (a,b), $c=1$ (c,d), $c=2$ (e,f), and $c=10$ (g,h).
The density of the asymptotic state is zero when two coordinates 
$\xi_i$ and $\xi_j$ ($i\neq j$) coincide.
}
\end{center}
\end{figure}
%

The connection between the asymptotic state (\ref{psiAsym}) and 
the initial state $\psi_{B0}$ is made through the Fourier transform 
of the initial fermionic wave function $\tilde \psi_F(\{ k \})$. 
More insight into the connection between the initial state 
and the asymptotic state can be made by expressing 
$\tilde \psi_F(\{ k \})$ through the coefficients 
$b(\{ k \})$ utilized in the expansion (\ref{LLcomplete}). 
First, let us note that the coefficients $b(\{ k \})=b(k_1,k_2,\ldots,k_N)$ 
are antisymmetric with respect to the interchange of any two arguments 
$k_i$ and $k_j$ $(i\neq j)$. 
This follows from the fact that the LL eigenstates 
$\psi_{\{ k \}}$ possess the same property, see Ref. \cite{Korepin1993}. 
By using this property of $b(\{ k \})$, Eq. (\ref{psiF0viab}) can 
be rewritten as 
\begin{align}
\psi_{F0} 
&={\mathcal N}_{c}^{-1} \sum_{\{ k \}} 
\mathcal{N}({\{ k \}}) b(\{ k \}) \sum_P (-)^P e^{i \sum_{j=1}^{N} k_{Pj}x_j}
\nonumber \\
&={\mathcal N}_{c}^{-1} \sum_P \sum_{\{ k \}} 
\mathcal{N}(k_{P1},k_{P2},\ldots,k_{PN}) 
\nonumber \\
& \times b(k_{P1},k_{P2},\ldots,k_{PN}) 
e^{i \sum_{j=1}^{N} k_{Pj}x_j}
\nonumber \\
&={\mathcal N}_{c}^{-1} N! \sum_{\{ k \}}
\mathcal{N}(\{ k \}) b(\{ k \}) e^{i \sum_{j=1}^{N} k_j x_j}.
\label{connection}
\end{align}
By comparing Eqs. (\ref{connection}) and (\ref{psiFt}) we obtain
\begin{align}
\tilde \psi_{F} (\{ k \})
&={\mathcal N}_{c}^{-1} N! \mathcal{N}(\{ k \}) b(\{ k \}).
\label{connection1}
\end{align}
Evidently, the Fourier transform of the initial fermionic wave function 
$\tilde \psi_{F} (\{ k \})$ is directly proportional to the projections 
$b(\{ k \})$ of the initial bosonic wave function onto the LL eigenstates. 
From this relation we can conclude that 
the asymptotic wave function (\ref{psiAsym}) has TG structure 
as a consequence of the antisymmetry of the coefficients $b(\{ k \})$,
which originates from the antisymmetry of the LL eigenstates 
with respect to $k_j$ arguments \cite{Korepin1993}. 
It is also worthy to note that Eq. (\ref{psiBt}), and therefore our main result, 
can be obtained without explicit use of the Fermi-Bose transformation; 
by writing the time dependent LL states as 
$\psi_{B,c} = \sum_{\{ k \}} b(\{ k \}) \psi_{\{ k \}} \exp(-i\sum_j k_j^2 t)$, 
and after employing the antisymmetry of $b(\{ k \})$ [equivalently 
as in Eq. (\ref{connection})] one obtains Eq. (\ref{psiBt}). 
Formulae (\ref{psiAsym}) and (\ref{connection1}) provide, under general
conditions, the asymptotic form of the wave functions for the 
freely expanding LL gas, and the connection between these asymptotic states 
and the initial states.

For the sake of the clarity of the paper, let us illustrate 
the asymptotic state of the LL gas on a particular example. 
Suppose that initially the LL gas is in the ground state, 
enclosed in an infinitely deep box of length $L$. The ground state 
$\psi_{B0}$ for this potential was found by employing the superposition of the 
Bethe ansatz wave functions in Ref. \cite{Gaudin1971}. 
The coefficients $b(\{ k \})$ can be relatively easily 
found for a few particles by employing a computer program for algebraic
manipulation (Mathematica).
In Fig. \ref{figdensity} we illustrate the initial state and the asymptotic state 
for the case of $N=3$ particles, and for values of $c=0.2,1,2$, and $10$,
by showing the contour plots of the probabilities $|\psi_{B0}(L/2,x_2,x_3)|^2$
(left column) and $|\psi_{\infty}(0,\xi_2,\xi_3)|^2$ (right column). 
Thus, one particle is fixed in the center of the system, while the plots illustrate the 
probability of finding the other two particles in space. 
The left column illustrating the initial states shows that the system 
becomes more correlated with increasing interaction strenght $c$
and it enters the TG regime for sufficiently large $c$ (e.g., 
for $c=10$ depicted in Fig. \ref{figdensity} (g) the ground state of the 
system is in the TG regime). 
The right column illustrating the asymptotic state shows that the 
wave function is zero whenever two of the coordinates coincide. 
However, it is important to note that the properties of the asymptotic
wave functions, even though they possess the TG structure, can significantly 
differ from the properties of the TG gas in the equilibrium ground state.
This can be seen by comparing the asymptotic state in Fig. \ref{figdensity} (b), 
and the TG ground state shown in Fig. \ref{figdensity} (g). 
The asymptotics of Fig. \ref{figdensity} (b) is obtained after free expansion from a 
weakly interacting ground state $(c=0.2)$; from Fig. \ref{figdensity} (b)
we observe that when one particle is fixed at zero, there is still a relatively 
large probability of finding the other two particles to the left and to the right 
of the fixed one. In contrast, for the TG ground state shown in 
Fig. \ref{figdensity} (g), if one particle is fixed in 
the center of the system, the other two 
are on the opposite sides of that one. 
Furthermore, by comparing the asymptotic states in Figs. \ref{figdensity}
(b), (d), (f), and (h), we see that their properties depend on the interaction 
strength $c$. 
It is worthy to mention again that free expansion can be utilized 
to deduce information on the initial state (see, e.g.,
Refs.~\cite{Bloch2008} and references therein); free 1D 
expansion can distinguish between different initial regimes of the 
LL gas \cite{Ohberg2002}.

Let us now address the case of attractive interactions.
For $c<0$, the cusp condition assumes a
form that is identical to that for $c>0$ (see, e.g., Ref.~\cite{Muga1998}).
Therefore, by acting on some fermionic time-dependent wave function
obeying Eq.~(\ref{SchF}) with the Fermi-Bose transformation
operator $\hat O_{c<0}$, one obtains an exact solution for the attractive
time-dependent LL gas in the form $\hat O_{c<0} \psi_F$ (see Appendix \ref{app:FB});
our derivation holds for this family of wave functions. 
Experiments where the attractive quasi-1D Bose gas is suddenly
released from a trapping potential were used to study
solitons made of attractively interacting BEC \cite{Khaykovich2002}.
Exact studies of such a system within the framework of the LL model
are expected to provide deeper insight into nonequilibrium phenomena
beyond the Gross-Pitaevskii mean-field regime, where interesting
dynamical effects can occur \cite{Streltsov2008,Buljan2005}.

It should be noted that the time scale it takes for the LL
system to reach the TG regime depends on the initial condition.
The next-to-leading term of the asymptotic wave function
is suppressed relative to the leading term by a factor $1/t$,
as obtained by the stationary phase expansion in Eq.~(\ref{psiBtSeries2}).
From this we can deduce the scaling of the interaction
energy, defined as
\begin{align}
E_\mathrm{int}
&=2c\int dx_1\cdots dx_N\, |\psi_{B,c}|^2
\sum_{1\leq i < j \leq N}\delta(x_i-x_j),
\label{intE}
\end{align}
as $t\rightarrow\infty$. Since the interaction
strength $c$ is finite, and since the asymptotic density
$|\psi_{\infty}(\xi_1,\ldots,\xi_N,t)|^2$ equals zero for any pair
of arguments being equal, $\xi_i=\xi_j$, one concludes that
asymptotically the leading term of the interaction energy vanishes.
Since the first correction to the leading TG
term of the wave function is of order $t^{-1}$, and since
$\delta(x_i-x_j)=t^{-1} \delta(\xi_i-\xi_j)$, the interaction
energy asymptotically decays to zero as
$E_\mathrm{int}\propto t^{-3}$. This power law decay of the interaction energy
is illustrated in the following section.

\section{Example: Fermionic wave function
expanding from a harmonic trap}
\label{Sec:Check}

In Ref.~\cite{Buljan2008}, we have constructed a particular family of time-dependent
wave functions describing a freely expanding LL gas.
The wave functions were obtained by acting with the Fermi-Bose mapping operator
onto a specific time-dependent fermionic wave function,
\begin{eqnarray}
\psi_F & \propto &
\exp\Big\{-i \frac{N^2 \nu}{2}\tau(t)
           - \frac{\nu-i \nu^2 t}{4}
           \sum_{j=1}^N \Big[\frac{x_j}{b(t)} \Big]^2 \Big\}
\nonumber \\
&& \times\ b(t)^{-{N^2}/{2}}
\prod_{1\leq i < j \leq N}(x_j-x_i),
\label{TDF}
\end{eqnarray}
which describes free expansion of noninteracting fermions in one spatial dimension.
The initial fermionic wave function at $t=0$ corresponds
to a fermionic ground state in a harmonic trap $V(x)=\nu^2 x^2/4$
(see, e.g., Ref.~\cite{Girardeau2001}).
Here, $\nu$ corresponds to the trapping frequency,
$b(t)=\sqrt{1+t^2\nu^2}$, and $\tau(t)=\arctan(\nu t)/\nu$.
The limiting form of the LL wave function,
$\psi_{B,c}(\eta_1 b(t),\ldots,\eta_N b(t),t)$,
for $t\rightarrow\infty$, was shown to have the following form characteristic
for a TG gas:
\begin{align}
&\psi_{B,c}(\eta_1 b(t),\ldots,\eta_N b(t),t)
\propto
b(t)^{-{N}/{2}}
\nonumber \\
&\quad\times\
\exp\Big\{-i \frac{N^2 \nu}{2}\tau(t)
          - \frac{\nu-i \nu^2 t}{4} \sum_{j=1}^N \eta_j^2 \Big\}
\nonumber\\
&\quad\times\
\prod_{1\leq i < j \leq N}g(\eta_j-\eta_i)+{\mathcal O}({1}/{t}),
\label{asymHO}
\end{align}
where $g(\eta)=|\eta|+i\nu \eta^2 / 2c$.
Equation (\ref{psiAsym}) is a generalization of this result given first in
Ref.~\cite{Buljan2008}.
%
\begin{figure}
\begin{center}
\includegraphics[width=0.45 \textwidth ]{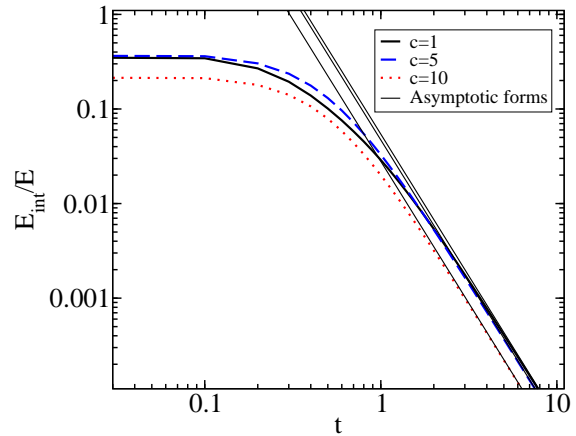}
\caption{ \label{Eintt}
(color online) Time-evolution of the interaction energy
$E_\mathrm{int}(t)$, expressed in units of the total energy $E$.
The three curves correspond
to values of $c=1$ (solid line), $c=5$ (dashed line),
and $c=10$ (dotted line). The straight lines depict the asymptotic
power law behavior of the interaction energy,
$E_\mathrm{int}(t)\propto t^{-3}$ (see text for details).
}
\end{center}
\end{figure}
%
Since Eq.~(\ref{psiAsym}) was obtained with the help of the stationary phase approximation,
whereas (\ref{asymHO}) is obtained straightforwardly from the exact form
of the specific LL wave function (see Ref.~\cite{Buljan2008}),
it is worthy to verify that Eq.~(\ref{psiAsym}) reproduces Eq.~(\ref{asymHO})
as a special case. In order to do so, we calculate the
Fourier transform of the initial fermionic wave function, i.e.,
$\psi_F(x_1,\ldots,x_N,0)$ from Eq.~(\ref{TDF}).
Interestingly, the Fourier transform has exactly the same functional
form as the initial condition in $x$-space:
\begin{equation}
\tilde \psi_F \propto
e^{-\sum_{j=1}^N k_j^2/\nu}
\prod_{1\leq i < j \leq N}(k_j-k_i).
\label{kTDF}
\end{equation}
By plugging this form into Eq.~(\ref{psiAsym}) we obtain:
\begin{eqnarray}
\psi_{\infty} & \propto & t^{-{N}/{2}}
e^{-\sum_{j=1}^N \xi_j^2/(4\nu)}
e^{({i}/{4})\sum_{j=1}^{N} \xi_j^2 t}
\nonumber \\
&&
\times\ \prod_{1\leq i<j\leq N}
[|\xi_j-\xi_i|+\frac{i}{2c}(\xi_j-\xi_i)^2].
\label{xiTDF}
\end{eqnarray}
After replacing $\xi_j=x_j/t$ with $\nu \eta_j=\nu x_j/b(t)$ which asymptotically
approaches $\nu \eta_j\sim x_j/t=\xi_j$,
we obtain the functional form identical to Eq.~(\ref{asymHO}).
This verifies the validity of Eq.~(\ref{psiAsym}) in the special
case studied in Ref.~\cite{Buljan2008}.

\begin{figure}
\begin{center}
\includegraphics[width=0.45 \textwidth ]{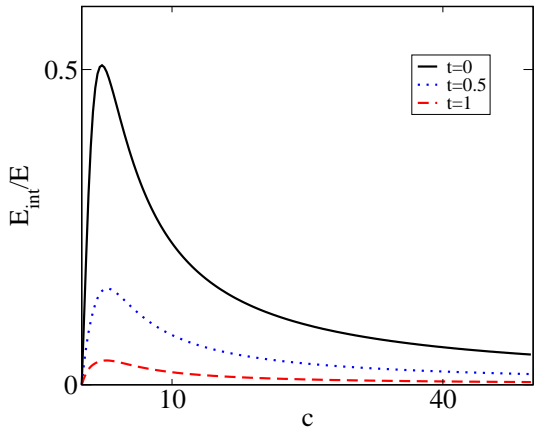}
\caption{ \label{Eintc}
(color online) The ratio $E_\mathrm{int}/E$ as a function of the
interaction strength $c$, at three values of time,
$t=0$ (solid line), $t=0.5$ (dotted line), and $t=1$ (dashed line)
(see text for details).}
\end{center}
\end{figure}

In order to verify the asymptotic power law decay of the interaction
energy $E_\mathrm{int}$ obtained in the previous section,
let us calculate the time-evolution
of $E_\mathrm{int}$ for the specific family of LL wave functions
discussed in this section. We calculate integral (\ref{intE}) for
$N=3$ particles, and $\nu=2$.
Given these parameters, $E_\mathrm{int}$ depends on the strength of
the interaction $c$ and time $t$. Figure \ref{Eintt} illustrates
time-evolution of the interaction energy for three values of $c$;
displayed curves depict the ratio $E_\mathrm{int}(t)/E$, where $E$ denotes
the total energy, which is a constant of motion.
Evidently, after some initial transient period the interaction energy starts
its asymptotic power law decay $E_\mathrm{int}(t)\propto t^{-3}$.
It should be noted that the contribution of the interaction energy
to the total energy depends on the interaction strength $c$.
This is illustrated in Fig. \ref{Eintc} which shows $E_\mathrm{int}(t)/E$
as a function of $c$ at three points in time.
At $t=0$, the contribution of the interaction energy to the total
energy is non-monotonous with the increase of $c$; it is zero at $c=0$ and
in the TG limit $c\rightarrow\infty$, with a specific maximal value
in between. The form of the curve is preserved
for finite values of $t$, with the evident decay of the interaction
energy to zero as $t\rightarrow\infty$.
Note that an equivalent non-monotonous behavior of the interaction energy
as a function of $c$ was found for the Lieb-Liniger gas in the ground
state for $c>0$ and with periodic boundary conditions \cite{Muga1998}.
%
%

\section{Asymptotic single-particle density}
\label{Sec:SPden}

Given the asymptotic form of the wave function,
we finally consider the asymptotic form of the single-particle density
which is of considerable interest for experiment.
The single-particle density is defined as
$\rho_c(x,t)=N\int dx_2\cdots dx_N |\psi_{B,c}(x,x_2,\ldots,x_N,t)|^2 $.
For studying asymptotics, it is convenient to define the
asymptotic form in terms of the rescaled coordinates $\xi=x/t$:
\begin{equation}
\rho_{\infty}(\xi)={\cal N}_{\infty} t^N
\int_{-\infty}^{\infty} d\xi_2\ldots d\xi_N
|\psi_{\infty}(\xi,\xi_2,\ldots,\xi_N,t)|^2;
\label{rhoInfty}
\end{equation}
here the normalization constant ${\cal N}_{\infty}$ is chosen such that
$\int d\xi\,\rho_{\infty}(\xi)=N$, the total number of particles,
while the factor $t^N$ cancels the trivial time-scaling of
the asymptotic single-particle density.

For the specific asymptotic form of the wave function
(\ref{xiTDF}) we can analytically calculate the asymptotic form
of the density for a few particles. As an example, for $N=3$,
the normalization constant is
\begin{equation}
{\cal N}_{\infty,N=3}=\frac{c^6}
{\sqrt{2\pi^{3} \nu^{9}}
(8 c^6 + 48 c^4 \nu + 90 c^2 \nu^2 + 45 \nu^3)},
\label{Ninfty}
\end{equation}
while the single-particle density has the following structure:
\begin{eqnarray}
\rho_{\infty}(\xi)
& = & {\cal N}_{\infty,N=3}
\frac{\pi\nu^2}{8 c^6} e^{-{ \xi^2 }/( 2 \nu)}
\label{rhoInfty3} \\
& &
\times\ [
32 c^6 (3 \nu^2 + \xi^4)
\nonumber \\
& &
+\ 16 c^4 (33 \nu^3 - 3 \nu^2 \xi^2 +
9 \nu \xi^4 + \xi^6)
\nonumber \\
& &
+\ 2 c^2 (465 \nu^4 - 60 \nu^3 \xi^2 + 90 \nu^2 \xi^4 +
20 \nu \xi^6 + \xi^8)
\nonumber \\
& &
+\ 3 \nu (165 \nu^4 - 60 \nu^3 \xi^2 + 30 \nu^2 \xi^4 +
4 \nu \xi^6 + \xi^8)].
\nonumber
\end{eqnarray}
This expression shows that the Gaussian shape of the
single-particle density is modulated with
the $N$-hump structure characteristic for the
single-particle density of a TG gas in the ground state
of some external potential.
The corresponding density (\ref{rhoInfty3}), in terms of
$\eta=\xi/\nu$ is shown in Fig. 2 of Ref.~\cite{Buljan2008}.
It should be noted that such an asymptotic form of the single-particle density
corresponds to a particular family of time-dependent wave functions 
obtained in Ref.~\cite{Buljan2008}.
For different initial conditions one can obtain a different shape 
of the asymptotic single-particle density as follows from Eqs.~(\ref{psiAsym})
and (\ref{connection1});
the asymptotic single-particle density depends on $\tilde \psi_F(\{ k \})$,
that is $b(\{ k \})$.

\section{Comparison with the hydrodynamic approximation}
\label{Sec:Hydro}

Besides providing insight into the physics of interacting 
time-dependent many-body systems, our motivation to study exact solutions 
of such systems is to utilize those solutions as a benchmark 
against various approximations. 
Free expansion of a Lieb-Liniger gas has been studied in Ref. \cite{Ohberg2002}
by employing the formalism introduced in Ref. \cite{Dunjko}, 
referred to as the hydrodynamic approximation. 
This formalism can be written in a form of a nonlinear 
evolution equation for a single-particle wave function 
$\psi_H(x,t)$ [see Eq. (9) in Ref. \cite{Ohberg2002}], 
\begin{equation}
i\frac{\partial \psi_H(x,t)}{\partial t}=
-\frac{\partial^2 \psi_H}{\partial x^2}+V(x)\psi_H+
c^2f \left (\frac{c}{|\psi_H|^2} \right)\psi_H,
\label{hydro}
\end{equation}
where $|\psi_H(x,t)|^2$ denotes the single-particle density 
normalized to $\int |\psi_H(x,t)|^2 dx=N$, while the function 
$f$ which appears in the nonlinear term is defined in Ref. 
\cite{Dunjko}, and also tabulated in Ref. [19] of Ref. \cite{Dunjko}. 
The potential is $V(x)=0$ during free expansion. 
The hydrodynamic approximation was used to obtain Eq. (\ref{hydro}), 
which is written in units corresponding to the Lieb-Liniger model of 
Eq. (\ref{LLmodel}).
The nonlinear equation above reduces to the standard Gross-Pitaevskii 
equation for small interactions, and to the nonlinear equation from 
Ref. \cite{Kolomeisky2000} for strong interactions \cite{Ohberg2002}. 
The hydrodynamic approximation overestimates the coherence
in the system, and therefore it may not be accurate for analyzing 
observables strongly connected to coherence. 
However, it is reasonable to compare the exact asymptotic form of the 
single-particle density after free expansion with the asymptotic
form obtained from the hydrodynamic approximation. 
%
\begin{figure}
\begin{center}
\includegraphics[width=0.45 \textwidth ]{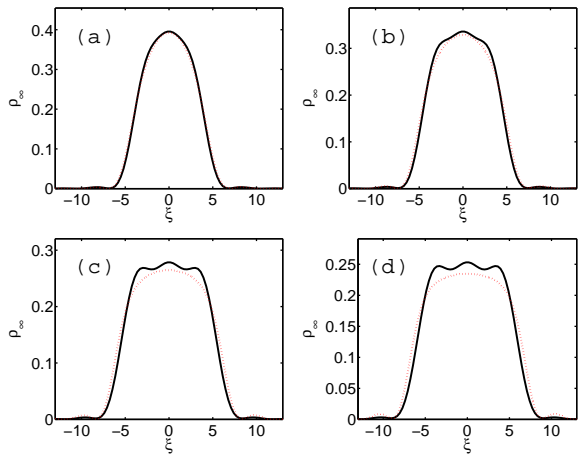}
\caption{ \label{comparison}
(color online) The asymptotic form of the SP density 
obtained exactly (black solid line), and with the hydrodynamic 
approach (red dotted line). The parameters used in the calculation 
are $N=3$, $L=\pi$, $c=1$ (a), $c=2$ (b), $c=5$ (c), and $c=10$ (d). 
(see text for details).}
\end{center}
\end{figure}
%

Let us follow upon our example from Section \ref{sec:ExpAsymp}, 
that is, let us consider the asymptotic form of the single particle density 
$\rho_{\infty}(\xi)$ of a LL gas which is initially in the ground state of a box 
with infinitely high walls; the length of the box is $L=\pi$. 
The calculation of the exact SP density demands performing 
multi-dimensional integration over $N-1$ variables 
which is not a simple task. For this reason, the number of particles 
in our calculation of the exact SP density is $N=3$. 
For the initial condition of the hydrodynamic approach $\psi_H(x,t=0)$ 
we could choose $\psi_H(x,t=0)=\sqrt{N/L}$ within the box, 
and zero otherwise. This would be a good initial condition 
in the thermodynamic limit (large $N$, $N/L=const.$). 
However, since for our exact calculation we used $N=3$, we have chosen, in
order to be able to compare between the two approaches, the hydrodynamic
initial field $\psi_H(x,t=0)=\sqrt{n_\mathrm{exact}}$, where
$n_\mathrm{exact}$ is the exact SP density of the initial ground state 
(this can be calculated by employing Ref. \cite{Gaudin1971}). 
Figure \ref{comparison} displays the exact asymptotic form of the SP density, 
and the hydrodynamic asymptotic SP density. The latter is obtained numerically 
by solving Eq. (\ref{hydro}) with the standard split-step Fourier technique;
the nonlinear term in Eq. (\ref{hydro}), that is, the function 
$f(c/|\psi_H(x)|^2)$ is calculated by using values tabulated in Ref. [19] of Ref. \cite{Dunjko}. 
The asymptotic dynamics in the hydrodynamic approach occurs
after sufficiently long propagation, when the SP density 
starts exhibiting self-similar propagation (see also \cite{Ohberg2002}). 

The agreement is qualitatively excellent for all values of the 
interaction strength, and quantitatively  excellent for $c<1$. 
The width of the SP density as a function of $\xi=x/t$ indicates 
the velocity of the expansion of the cloud. 
The asymptotic FWHM (full-width at half maximum) expansion velocity is 
in good agreement for all values of $c$. 
The hydrodynamic approximation does not reproduce 
small humps in the SP density, characteristic in the TG regime 
after expansion from the ground state; this discrepancy 
is expected to be smaller if we had calculated expansion
from the ground state with large $N$, where the hydrodynamic approximation 
is expected to work even better. 

Another possible comparison that can be made with the hydrodynamic
approximation is the following. The LL wave function which is utilized as the 
initial condition in Sec. \ref{Sec:Check} and Ref. \cite{Buljan2008} is obtained
by acting with the operator $\hat O_c$ onto the fermionic ground state $\psi_{F0}$
in the harmonic trapping potential $V(x)=\nu^2x^2/4$. 
This wave function can approximate the ground state 
only when the commutator $[\hat O_c,V(x)]$ can be neglected \cite{Buljan2008}. 
The SP density of this state can be compared with the static hydrodynamic 
density obtained in Ref. \cite{Dunjko} for the LL gas in a harmonic trap. 
Due to the properties of the operator $\hat O_c$ \cite{Buljan2008} 
and the fermionic ground state in the harmonic trap $\psi_{F0}$, 
it is straightforward to verify that the shape of the SP density
corresponding to the state $\hat O_c \psi_{F0}$ scales
as $\rho(x)\rightarrow \rho(x/s)/s$ under the transformation 
$\nu\rightarrow \nu/s^2$, $c\rightarrow c/s$, that is, the shape of the 
SP density does not change under this transformation. 
The same is true for
the shape of the (ground-state) SP density obtained with
the hydrodynamic approach, which has been shown \cite{Dunjko} to
depend on a single parameter $\eta = (\frac{3N\nu}{4c^2})^\frac{2}{3}$ that is invariant under the 
transformation $\nu\rightarrow \nu/s^2$, $c\rightarrow c/s$. 
This is fully analogous to the case of a homogeneous LL gas
where the only governing parameter $\gamma = c/n$ is invariant
under a simultaneous rescaling of the interaction strength
$c$ and the linear particle density $n$ \cite{Lieb1963}. 
The shape of the SP density of the state 
$\hat O_c \psi_{F0}$ (calculated for $N=3$) agrees with the shape 
obtained in Ref. \cite{Dunjko} only in the Tonks-Girardeau limit ($\eta\ll 1$) 
where $\hat O_c \psi_{F0}$ is a good approximation for the ground state. 
If we reduce the interaction strength $c$ by keeping $\nu$ fixed, 
thereby increasing $\eta$, the two SP densities will no longer have 
a similar shape; this stems from a simple fact that $\hat O_c \psi_{F0}$ is 
an excited state for sufficiently small values of $c$, 
because the commutator $[\hat O_c,V(x)]$ cannot be neglected, 
whereas the hydrodynamic solution approximates the ground state.

\section{Conclusion}

We have derived the asymptotic form of the
wave function describing a freely expanding Lieb-Liniger gas.
It is shown to have Tonks-Girardeau structure [see Eq.~(\ref{psiAsym})],
that is, the wave functions vanish when any two of the particle coordinates coincide. 
We have pointed out that the properties of these asymptotic states 
can significantly differ from the properties of a TG gas 
in a ground state of an external potential [see Fig.~\ref{figdensity}]. 
The dependence of the asymptotic state on the initial state 
is discussed [see Eq.~(\ref{connection1})]. 
The analysis was performed for time-dependent Lieb-Liniger wave functions 
which can be obtained through the Fermi-Bose transformation (\ref{ansatz}). 
This encompasses initial conditions which correspond to the ground state 
of a repulsive Lieb-Liniger gas in physically realistic external potentials. 
Thus, our analysis characterizes the free expansion from such a ground
state, after the potential is suddenly switched off.
In deriving our main result, Eq.~(\ref{psiAsym}), we have used
the stationary phase approximation. This generalizes and adds upon the result
from Ref.~\cite{Buljan2008} which was derived for a particular
family of time-dependent Lieb-Liniger wave functions.
We have demonstrated that the interaction energy
of the freely expanding LL gas asymptotically decays according to a power law,
$E_\mathrm{int}\propto t^{-3}$. 
Furthermore, we have calculated the asymptotic single-particle
density for free expansion of a LL gas from an infinitely deep box 
potential. We have compared our exact calculation with the hydrodynamic approximation 
introduced in Ref. \cite{Dunjko}, and employed in Ref. \cite{Ohberg2002}
in the context of free expansion, obtaining good agreement 
for all values of the interaction strength. As a possible future avenue of research, 
we point out that the methodology employed here for the analysis 
of asymptotic wave functions has the potential to be 
exploited further to study the evolution of various observables (e.g., 
the momentum distribution which was studied for a TG gas \cite{Rigol2005}) 
and correlations (e.g., see \cite{AMRey2008} and Refs. therein) during free 
expansion.


\acknowledgments
We are grateful to M.~Fleischhauer and V. Dunjko for very useful comments
and suggestions. H.B. and R.P. acknowledge support by the Croatian Ministry of
Science (MZO\v S) (Grant No.~119-0000000-1015).
T.G. acknowledges support by the Deutsche Forschungsgemeinschaft.
This work is also supported by the Croatian-German
scientific collaboration funded by DAAD and MZO\v S, and 
in part by the National Science Foundation under
Grant No. PHY05-51164. 
\\[-4ex]

\begin{appendix}
\section{Fermi-Bose transformation}
\label{app:FB}
In this appendix we
outline the proof that
the wave function (\ref{ansatz}) obeys both
the cusp condition imposed by the interactions and
Eq.~(\ref{free}), i.e., that it obeys Eq. (\ref{LLmodel}).
Without loss of generality we restrict our discussion to
the fundamental permutation sector $R_1$.
Let us write the differential operator as
$\hat O_c=\prod_{1\leq i < j \leq N} \hat B_{ij}$,
where
\begin{equation}
\hat B_{ij}=\left[
1+\frac{1}{c}
\left(
\frac{\partial}{\partial x_{j}}-
\frac{\partial}{\partial x_{i}}
\right)
\right]
\label{Bij}.
\end{equation}
We first show that the wave function (\ref{ansatz})
obeys the cusp condition (\ref{interactions}) (see Ref.~\cite{Korepin1993}).
Consider an auxiliary wave function
\begin{align}
\psi_\mathrm{AUX}(x_1,\ldots,x_N,t)
& = \hat B_{j+1,j} \hat O_c \psi_F \nonumber \\
& =  \hat B_{j+1,j} \hat B_{j,j+1}\hat O'_{j,j+1} \psi_F,
\end{align}
where the primed operator $\hat O'_{j,j+1}=\hat O_c/\hat B_{j,j+1}$ omits
the factor $\hat B_{j,j+1}$ as compared to $\hat O_c$.
The auxiliary function can be written as
\begin{equation}
\psi_\mathrm{AUX}= \left [
1-\frac{1}{c^2}
\left (
\frac{\partial}{\partial x_{j+1}}-\frac{\partial}{\partial x_j}
\right)^2
\right] \hat O'_{j,j+1} \psi_F.
\label{AuxAsym}
\end{equation}
It is straightforward to verify that the operator
$\hat B_{j+1,j} \hat B_{j,j+1}\hat O'_{j,j+1}$ in front of $\psi_F$ is
invariant under the exchange of $x_j$ and $x_{j+1}$.
On the other hand, the fermionic wave function $\psi_F$ is fully
antisymmetric with respect to the interchange of $x_j$ and $x_{j+1}$.
Thus, $\psi_\mathrm{AUX}(x_1,\ldots,x_j,x_{j+1},\ldots,x_N,t)$ is antisymmetric
with respect to the interchange of $x_j$ and $x_{j+1}$,
which leads to
\begin{equation}
\psi_\mathrm{AUX}(x_1,\ldots,x_j,x_{j+1},\ldots,x_N,t)|_{x_{j+1}=x_{j}}=0.
\end{equation}
This is fully equivalent to the cusp condition (\ref{interactions}),
$\hat B_{j+1,j} \psi_{B,c}|_{x_{j+1}=x_{j}}=0$.
Thus, the wave function (\ref{ansatz}) obeys constraint
(\ref{interactions}) by construction.

Second, from the commutators $[\partial^2/\partial x_i^2,\hat O_c]=0$ and
$[i\partial/\partial t,\hat O_c]=0$ follows that if $\psi_F$
obeys Eq.~(\ref{SchF}), then $\psi_{B,c}$ obeys
Eq.~(\ref{free}), which completes the proof.

If we use the expression
\begin{equation}
\hat B_{ij}=\left[
\mathrm{sgn}(x_j-x_i)+\frac{1}{c}
\left(
\frac{\partial}{\partial x_{j}}-
\frac{\partial}{\partial x_{i}}
\right)
\right],
\label{Bijsgn}
\end{equation}
we obtain $\hat O_c=\prod_{1\leq i < j \leq N} \hat B_{ij}$
as in Eq. (\ref{oO}), which is valid inside any sector of the
configuration space (see \cite{Gaudin1983}).
Note that for $c\rightarrow\infty$,
one recovers Girardeau's Fermi-Bose mapping \cite{Girardeau1960},
where the operator $\hat O_{c=\infty}=\prod_{1\leq i < j \leq N} \mbox{sgn}(x_j-x_i)$
maps a noninteracting fermionic to a bosonic
Tonks-Girardeau wave function.
\end{appendix}



\begin{thebibliography}{99}

\bibitem{Lieb1963}
E. Lieb and W. Liniger,
Phys. Rev. {\bf 130}, 1605 (1963);
\newline
E. Lieb, Phys. Rev. {\bf 130}, 1616 (1963).

\bibitem{Girardeau1960}
M. Girardeau,
J. Math. Phys. {\bf 1}, 516 (1960).

\bibitem{OneD}
F. Schreck, L. Khaykovich, K.L. Corwin, G. Ferrari, T. Bourdel,
J. Cubizolles, and C. Salomon,
Phys. Rev. Lett. {\bf 87}, 080403 (2001);
A. G\" orlitz, J.M. Vogels, A.E. Leanhardt, C. Raman, T.L. Gustavson,
J.R. Abo-Shaeer, A.P. Chikkatur, S. Gupta, S. Inouye, T. Rosenband,
and W. Ketterle, {\em ibid.} {\bf 87}, 130402  (2001);
M. Greiner, I. Bloch, O. Mandel, T.W. Hansch, and T. Esslinger,
{\em ibid.} {\bf 87}, 160405 (2001);
H. Moritz, T. St\" oferle, M. Kohl, and T. Esslinger,
{\em ibid.} {\bf 91}, 250402 (2003);
B. Laburthe-Tolra, K.M. O'Hara, J.H. Huckans, W.D. Phillips, S.L. Rolston,
and J.V. Porto, {\em ibid.} {\bf 92}, 190401 (2004);
T. St\" oferle, H. Moritz, C. Schori, M. Kohl, and T. Esslinger,
{\em ibid.} {\bf 92},  130403 (2004).

\bibitem{TG2004}
T. Kinoshita, T. Wenger, and D.S. Weiss,
Science {\bf 305}, 1125 (2004);
B. Paredes, A. Widera, V. Murg, O. Mandel, S. F\" olling,
I. Cirac, G. V. Shlyapnikov, T. W. H\" ansch, and I. Bloch,
Nature (London) {\bf 429}, 277 (2004).

\bibitem{Kinoshita2006}
T. Kinoshita, T. Wenger, and D.S. Weiss,
Nature (London) {\bf 440}, 900 (2006).

\bibitem{Olshanii}
M. Olshanii, Phys. Rev. Lett. {\bf 81}, 938 (1998).

\bibitem{Petrov}
D.S. Petrov, G.V. Shlyapnikov, and J.T.M. Walraven,
Phys. Rev. Lett. {\bf 85} 3745 (2000).

\bibitem{Dunjko}
V. Dunjko, V. Lorent, and M. Olshanii,
Phys. Rev. Lett. {\bf 86} 5413 (2001).


\bibitem{Gaudin1983}
M. Gaudin, \textit{La fonction d'Onde de Bethe} (Paris, Masson, 1983).

\bibitem{Girardeau2003}
M.D. Girardeau,
Phys. Rev. Lett. {\bf 91}, 040401 (2003).

\bibitem{Buljan2008}
H. Buljan, R. Pezer, and T. Gasenzer,
Phys. Rev. Lett. {\bf 100}, 080406 (2008).


\bibitem{Girardeau2000}
M.D. Girardeau and E.M. Wright,
Phys. Rev. Lett. {\bf 84}, 5691 (2000).

\bibitem{Girardeau2000a}
M.D. Girardeau and E.M. Wright,
Phys. Rev. Lett. {\bf 84} 5239 (2000).

\bibitem{Ohberg2002}
P. \" Ohberg and L. Santos,
Phys. Rev. Lett. {\bf 89}, 240402 (2002);
P. Pedri, L. Santos, P. \" Ohberg, and S. Stringari,
Phys. Rev. A {\bf 68}, 043601 (2003). 

\bibitem{Busch2003}
T. Busch and G. Huyet, 
J. Phys. B {\bf 36} 2553 (2003). 

\bibitem{Rigol2005}
M. Rigol and A. Muramatsu,
Phys. Rev. Lett. {\bf 94}, 240403 (2005);
{\em ibid.} Mod. Phys. Lett. B {\bf 19}, 861 (2005).

\bibitem{Minguzzi2005}
A. Minguzzi and D.M. Gangardt,
Phys. Rev. Lett. {\bf 94}, 240404 (2005).

\bibitem{DelCampo2006}
A. del Campo and J.G. Muga,
Europhys. Lett. {\bf 74}, 965 (2006).

\bibitem{Rigol2006}
M. Rigol, V. Dunjko, V, Yurovskii, and M. Olshanii,
Phys. Rev. Lett. {\bf 98}, 050405 (2007).

\bibitem{Buljan2006}
H. Buljan, O. Manela, R. Pezer, A. Vardi, and M. Segev,
Phys. Rev. A {\bf 74}, 043610 (2006).

\bibitem{Pezer2007}
R. Pezer and H. Buljan,
Phys. Rev. Lett. {\bf 98}, 240403 (2007).

\bibitem{Gangardt2007}
D.M. Gangardt and M. Pustilnik,
Phys. Rev. A {\bf 77}, 041604 (2008).


\bibitem{McGuire1964}
J.B. McGuire,
J. Math Phys. (NY) {\bf 5}, 622, (1964).

\bibitem{Gaudin1971}
M. Gaudin,
Phys. Rev. A {\bf 4}, 386 (1971).

\bibitem{Muga1998}
J.G. Muga and R.F. Snider,
Phys. Rev. A {\bf 57}, 3317  (1998).

\bibitem{Sakmann2005}
K. Sakmann, A.I. Streltsov, O.E. Alon, and L.S. Cederbaum,
Phys. Rev. A {\bf 72}, 033613 (2005).

\bibitem{Batchelor2005}
M.T. Batchelor, X.-W. Guan, N. Oelkers, and C. Lee,
J. Phys. A {\bf 38}, 7787 (2005).

\bibitem{Kinezi2006}
Y. Hao, Y. Zhang, J.Q. Liang, and S. Chen,
Phys. Rev. A {\bf 73}, 063617 (2006).

\bibitem{Sykes2007}
A.D. Sykes, P.D. Drummond, and M.J. Davis,
Phys. Rev. A {\bf 76}, 063620 (2007).

\bibitem{Kanamoto2005}
R. Kanamoto, H. Saito, and M. Ueda,
Phys. Rev. Lett. {\bf 94}, 090404 (2005).

\bibitem{Creamer1981}
D.B. Creamer, H.B. Thacker, and D. Wilkinson,
Phys. Rev. D {\bf 23}, 3081 (1981).

\bibitem{Jimbo1981}
M. Jimbo and T. Miwa, Phys.
Rev. D {\bf 24}, 3169 (1981).

\bibitem{Korepin1993}
V.E. Korepin, N.M. Bogoliubov, and A.G. Izergin,
\textit{Quantum Inverse Scattering Method and Correlation Functions}
(Cambridge, Cambridge University Press, 1993).

\bibitem{Kojima1997}
T. Kojima, V.E. Korepin, N.A. Slavnov,
Commun. Math. Phys. {\bf 188}, 657 (1997)

\bibitem{Olshanii2003}
M. Olshanii and V. Dunjko,
Phys. Rev. Lett. {\bf 91}, 090401 (2003).

\bibitem{Gangardt2003}
D.M. Gangardt and G.V. Shlyapnikov,
New J. of Phys. {\bf 5}, 79 (2003).

\bibitem{Astrakharchik2003}
G.E. Astrakharchik and S. Giorgini,
Phys. Rev. A {\bf 68}, 031602(R) (2003).

\bibitem{Kheruntsyan2005}
K.V. Kheruntsyan, D.M. Gangardt, P.D. Drummond, and G.V. Shlyapnikov,
Phys. Rev. A {\bf 71}, 053615 (2005).

\bibitem{Forrester2006}
P.J. Forrester, N.E. Frankel, and M.I. Makin,
Phys. Rev. A {\bf 74}, 043614 (2006).

\bibitem{Caux2007}
J.-S. Caux, P. Calabrese, and N. A. Slavnov,
J. Stat. Mech. (2007) P01008.

\bibitem{Calabrese2007}
P. Calabrese and J.-S. Caux,
Phys. Rev. Lett. {\bf 98}, 150403 (2007).

\bibitem{Khodas2007}
M. Khodas, M. Pustilnik, A. Kamenev, and L.I. Glazman,
Phys. Rev. Lett. {\bf 99}, 109405 (2007).

\bibitem{Berman2004}
G.P. Berman, F. Borgonovi, F.M. Izrailev, and A. Smerzi,
Phys. Rev. Lett. {\bf 92}, 030404 (2004).

\bibitem{Li2005}
W. Li, X. Xie, Z. Zhan, and X. Yang,
Phys. Rev. A {\bf 72}, 043615 (2005).

\bibitem{Das2002}
K.K. Das, M.D. Girardeau, and E.M. Wright,
Phys. Rev. Lett. {\bf 89}, 170404 (2002).

\bibitem{Cheon1999}
T. Cheon and T. Shigehara,
Phys. Rev. Lett. {\bf 82}, 2536 (1999).

\bibitem{Yukalov2005}
V.I. Yukalov and M.D. Girardeau,
Laser Phys. Lett {\bf 2}, 375 (2005).

\bibitem{Bloch2008}
I. Bloch, J. Dalibard, and W. Zwerger,
arXiv: 0704.3011v2 (2007).

\bibitem{Calabrese2007a}
P. Calabrese and J. Cardy,
J. Stat. Mech. P06008 (2007).

\bibitem{Rigol2007a}
M. Rigol, V. Dunjko, and M. Olshanii,
Nature (London) 452, 854 (2008).

\bibitem{Dorlas1993}
T.C. Dorlas,
Commun. Math. Phys. {\bf 154} 347 (1993).

\bibitem{Khaykovich2002}
L. Khaykovich,  F. Schreck,  G. Ferrari,  T. Bourdel,  J. Cubizolles,
L. D. Carr,  Y. Castin,  C. Salomon,
Science {\bf 296} 1290 (2002).

\bibitem{Buljan2005}
H. Buljan, M. Segev, and A. Vardi,
Phys. Rev. Lett. {\bf 95} 180401 (2005).

\bibitem{Streltsov2008}
A.I. Streltsov, O.E. Alon, and L.S. Cederbaum,
Phys. Rev. Lett. {\bf 100} 130401 (2008).

\bibitem{Girardeau2001}
M.D. Girardeau, E.M. Wright, and J.M. Triscari,
Phys. Rev. A {\bf 63}, 033601 (2001).

\bibitem{Kolomeisky2000}
E.B. Kolomeisky, T.J. Newman, J.P. Straley, and X. Qi,
Phys. Rev. Lett. {\bf 85}, 1146 (2000).

\bibitem{AMRey2008}
E. Toth, A.M. Rey, R.P. Blakie, 
arXiv:0803.2922 (2008). 


\end{thebibliography}
\end{document}